\documentclass{aa}  

\usepackage{graphicx}
\usepackage{txfonts}
\usepackage{natbib} 
\bibpunct{(}{)}{;}{a}{}{,} 

\usepackage{verbatim} 
\usepackage{hyperref}   
\hypersetup{colorlinks=true,linkcolor=blue,citecolor=blue,filecolor=blue,urlcolor=blue}

\begin{document}

   \title{Stellar model calibrations with the Ai Phe binary system} 

   \subtitle{Open questions about the  robustness of the fit}
  \author{G. Valle \inst{1, 2}, M. Dell'Omodarme \inst{1}, P.G. Prada Moroni
        \inst{1,2}, S. Degl'Innocenti \inst{1,2} 
}
\titlerunning{Calibration with Ai Phe binary system}
\authorrunning{Valle, G. et al.}

\institute{
        Dipartimento di Fisica "Enrico Fermi'',
        Universit\`a di Pisa, Largo Pontecorvo 3, I-56127, Pisa, Italy
        \and
        INFN,
        Sezione di Pisa, Largo Pontecorvo 3, I-56127, Pisa, Italy
}

   \offprints{G. Valle, valle@df.unipi.it}

   \date{Received ; accepted }

  \abstract
{}
{Relying on recently available  and  very precise observational data for the Ai Phe binary system, we explore the robustness of the calibration of stellar models achievable with this system. }
{We adopt the SCEPtER pipeline with a fitting grid of stellar models computed for different initial chemical compositions and convective core overshooting efficiencies. We investigated the impact of different assumptions about the surface efficiency of microscopic diffusion, whose efficiency is still debated in the mass range of the system. We obtained the fit of this system adopting two alternative scenarios. In the reference scenario, we allowed modification of the surface metallicity due to microscopic diffusion, while in the  alternative scenario we assumed that competing mixing from other sources cancels out this effect.}
{Due to the fact that the primary star has already experienced the first dredge-up while the secondary has not, the tested scenarios show interesting differences. While the estimated age is quite robust, changing from $4.70^{+0.13}_{-0.14}$ Gyr to $4.62^{+0.13}_{-0.06}$ Gyr, the calibration of the convective core overshooting parameter $\beta$ reveals noticeable differences. The reference scenario suggests a wide multi-modal range of possible values of $\beta$, peaking around 0.10; on the contrary the alternative scenario computations point towards a sharp and lower $\beta$, peaking around 0.04.  }
{The impossibility to obtain an unambiguous fit confirms the difficulty in achieving a sensible calibration of the  free parameters of stellar models using binary systems, even when very accurate masses and radii are available. The results also suggest that the biases due to the assumptions underlying the stellar track computations may be different from one binary system to another.   } 

   \maketitle

\section{Introduction}\label{sec:intro}

Detached double-lined eclipsing binaries  allow precise and accurate measurements of the masses, radii, and effective temperatures of the components, and it is reasonable to assume that the system components share a common age and initial chemical composition. Therefore, these systems are routinely adopted as excellent experimental environments in which to test models of stellar evolution and structure in order to gain a deeper understanding of some physical processes, such as the treatment of convection or diffusion \citep[see among many][]{Andersen1991, Torres2010,  TZFor, Claret2017, CPD2023}.
Many authors have attempted to determine the exact relation, if any, between the stellar mass and the value of the convective core overshooting parameter, with opposing findings \citep[see][for a review]{Anders2023}. 
It is widely recognised that only systems observed with outstanding precision can provide further insight into this topic \citep[see e.g.][]{TZFor, Miller2020, Helminiak2021, Anders2023}.  

A perfect target for this investigation is AI Phoenicis (AI Phe, HD 6980), an eclipsing binary system composed of two stars with masses of around 1.2 $M_{\sun}$, for which very high-precision masses and radii are available.  
Several works in the literature propose an age estimate for this system \citep[e.g.][]{Andersen1988, Ribas2000, Kirkby-Kent2016}.
Recently, observations by \citet{Miller2020} significantly improved the precision  of the estimated effective temperatures for the system. It is therefore interesting to investigate how these new observations impact the fit of the system.
In the present paper, we attempt to calibrate the age and the convective core overshooting efficiency whilst assisted by this new data set.  Besides the obvious interest in obtaining these estimates, we are particularly interested in the exploration of possible systematic effects that may undermine the robustness of the obtained calibrations.   

{The Ai Phe system is composed of a primary in the early red giant branch (RGB) and a secondary in the early subgiant branch (SGB) phase. 
The stellar masses of the two objects are in a range where the convective envelope of both stars nearly vanishes in the main sequence (MS).
The lack of such an envelope may lead to noticeable variations of the surface chemical abundances during the MS evolution owing to different mixing processes. 
Many stellar models adopted for binary system studies account for the effect of microscopic diffusion during the stellar evolution (e.g. PARSEC, MIST, DSEP,  GARSTEC, or Pisa  models \citealt{PARSEC2022, Choi2016, Dotter2008, Weiss2008, database2012}), which has the potential to significantly modify the surface chemical composition and to change the internal characteristics in a non-negligible way. Indeed, including the effect of microscopic diffusion in stellar model computations has been shown to be of fundamental importance for correctly predicting the internal structure of the Sun \citep[see e.g.][]{Bahcall2001, Dalsgaard2007}, while the diffusion efficiency in Galactic GC stars is still debated \citep[see e.g.][]{Korn2007,  Gratton2011, Nordlander2012, Gruyters2014}. 
The adoption of unmoderated microscopic diffusion results in a steep drop of the surface [Fe/H] in the MS phase, a drop that nearly cancels out in the first part of the RGB when the external convection sinks down to regions in which helium and heavy elements previously diffused (first dredge-up). Microscopic diffusion is not the only mixing mechanism able to modify the surface chemical abundances, and other competing mechanisms have been investigated, such as rotational mixing, turbulence, mass advection, and radiative acceleration  \citep[e.g.][]{Eggenberger2010, Vick2010, Deal2020, Dumont2021}. However, the effort of including the effects of mechanisms
competing with microscopic diffusion in the stellar model computations  in a physically consistent way is still ongoing and will require considerable theoretical progress \citep[see e.g.][]{Moedas2022}.

As a consequence of these theoretical difficulties, a firm prediction of the surface metallicity for stars in the mass range of Ai Phe before the first dredge-up is problematic. This poses an interesting question as to the robustness of the fit obtained for similar systems, because knowledge of the surface metallicity is recognised to be of utmost importance in order to break the age--metallicity degeneracy and obtain any meaningful calibration from binary systems \citep[see e.g.][]{Lastennet2002, Torres2010, Higl2017}. 
Different stellar codes, relying on different assumptions about the microscopic diffusion efficiency, and given the inclusion of extra-mixing mechanisms, may obtain different age and/or overshooting efficiency calibrations on the same target system. Given the constant and progressive refinement of the observational precision, quantifying this potential bias is extremely relevant.

The structure of the paper is as follows. In Sect.~\ref{sec:method}, we 
discuss the method and the grids used in the estimation process. 
The  result of the calibration is
 presented in Sect.~\ref{sec:results-AiPhe} with an analysis of its robustness and comparison with the literature in Sect.~\ref{sec:discuss}.
Some concluding remarks are provided in Sect.~\ref{sec:conclusions}.

\section{Methods and observational constraints}\label{sec:method}

\subsection{Fitting technique}

The analysis is conducted adopting the SCEPtER pipeline\footnote{Publicly available on CRAN: \url{http://CRAN.R-project.org/package=SCEPtER}, \url{http://CRAN.R-project.org/package=SCEPtERbinary}}, a well-tested technique for fitting single and binary systems
\citep[e.g.][]{scepter1,eta,bulge, binary, TZFor}. 
The pipeline estimates  the parameters of interest (i.e. the system age, its initial chemical abundances, and the core overshooting parameter) adopting a grid maximum likelihood  approach.

The method we use is explained in detail in \citet{binary}; here, we provide only a brief summary for convenience. For every $j$-th point in the fitting grid of precomputed stellar models, a likelihood estimate is obtained for both stars:
\begin{equation}
        {{\cal L}^{1,2}}_j = \left( \prod_{i=1}^n \frac{1}{\sqrt{2 \pi}
                \sigma_i} \right) 
        \times \exp \left( -\frac{\chi^2}{2} \right)
        \label{eq:lik}
        ,\end{equation}
\begin{equation}
        \chi^2 = \sum_{i=1}^n \left( \frac{o_i -
                g_i^j}{\sigma_i} \right)^2
        \label{eq:chi2},
\end{equation}
where $o_i$ are the $n$ observational constraints, $g_i^j$ are the $j$-th grid point corresponding values, and $\sigma_i$ are the observational uncertainties.

The joint likelihood of the system is then computed as the product of the single star likelihood functions.  
It is possible to obtain estimates both for the individual components and for 
the  whole system. In the former case, the fits for the two stars are obtained independently,
while in the latter case the two objects must have a common age (with a tolerance of 1 Myr), identical initial helium abundance, and initial metallicity.

The error on the estimated parameters is obtained by means of Monte Carlo simulations. 
We generate $N = 10\,000$ artificial binary systems, sampling from a multivariate Gaussian distribution centred on the observational data, taking into account the correlation structure among the observational data for the two stars. 
As in \citet{TZFor}, we assume a correlation of 0.95 between
the primary and secondary effective temperatures, and 0.95
between the metallicities of the two stars. Regarding mass and radius correlations, the high precision of the estimates means that these parameters are of no importance, but we set them to 0.8 for the mass and -0.9 for the radius, which are typical values for this class of stars \citep{binary, TZFor}. Different correlation values for mass and radius  lead to negligible modifications of the results. 

\subsection{Observational data}

\begin{table}
        \centering
        \caption{Masses, radii, effective temperatures, and surface metallicities adopted as observational constraints in the fit of the Ai Phe binary system.}
        \label{tab:input}
        \begin{tabular}{lcc}
                \hline\hline
                & primary & secondary \\
                \hline 
                $M$ ($M_{\sun}$) & $1.2438 \pm 0.0008$ & $1.1938 \pm 0.0008$ \\
                $R$ ($R_{\sun}$) & $2.9303 \pm 0.0023$ & $1.8036 \pm 0.0022$ \\
                $T_{\rm eff}$ (K) &  $5094 \pm 50$ &  $6199 \pm 50$ \\
                ${\rm [Fe/H]}$ & $-0.14 \pm 0.1$ & $-0.14 \pm 0.1$\\
                \hline
        \end{tabular}
\end{table}

As observational constraints,  we use the masses, radii, metallicities [Fe/H], and effective temperatures of both stars. The adopted values and their uncertainties reported in Table~\ref{tab:input} are taken from \citet{Miller2020}.

The uncertainties in the effective temperature reported in \citet{Miller2020} are 16 K and 22 K for the primary and secondary component, respectively. These values do not account for the systematic effects that might modify the calibration scale, which were quantified in that paper as 11 K. 
In light of the existing difference in the calibration scale among different literature sources, we adopt a conservative approach to this observational constraint, assuming an uncertainty of 50 K for both stars.

\subsection{Stellar models grid}

The grids of models were computed for the exact masses of the target stars, from the pre-main sequence up to the start of the RGB or to the RGB tip for the more evolved, primary star in the Ai Phe system. 
The initial metallicity [Fe/H] was varied from $-0.4$ dex to 0.3 dex with
a step of 0.01 dex. 
We adopted the solar heavy-element mixture by \citet{AGSS09}; a test conducted adopting the \citet{GS98} heavy-element mixture showed negligible differences in the results. 
Several initial helium abundances were considered at fixed metallicity by adopting the commonly used
linear relation $Y = Y_p+\frac{\Delta Y}{\Delta Z} Z$
with the primordial abundance of  $Y_p = 0.2471$ from \citet{Planck2020}.
The helium-to-metal enrichment ratio $\Delta Y/\Delta Z$ was varied
from 1.0 to 3.0 with a step of 0.1 \citep{gennaro10}. 

Models were computed with the FRANEC code, in the same
configuration as was adopted to compute the Pisa Stellar
Evolution Data Base\footnote{\url{http://astro.df.unipi.it/stellar-models/}} 
for low-mass stars \citep{database2012}. 
The models were computed
assuming the solar-scaled mixing-length parameter $\alpha_{\rm
        ml} = 1.74$.
The extension of the extra-mixing region beyond the Schwarzschild border
was considered only for the primary star and was parametrised  in terms of the pressure scale height $H_{\rm 
        p}$: $l_{\rm ov} = \beta H_{\rm p}$, with 
$\beta$ in the range
[0.00; 0.28] with a step of 0.005. The code adopts step overshooting assuming an instantaneous mixing in the overshooting treatment. The radiative temperature gradient is adopted in the overshooting  region \citep[see][for more details of the overshooting implementation]{scilla2008}. 
Atomic diffusion was included adopting the coefficients given by
\citet{thoul94} for gravitational settling and thermal diffusion. 
To prevent extreme variations in the surface chemical abundances for stars
without a convective envelope, a diffusion inhibition mechanism similar to the one discussed in \citet{Chaboyer2001} was adopted.
The diffusion velocities were
multiplied by a suppression parabolic factor that takes a value of 1 for 99\% of the mass of the structure and 0 at the base of the atmosphere.
Further details of the stellar models are fully described in \citet{eta,binary} and references therein.  

Raw stellar evolutionary tracks were reduced to a set of tracks with the same number of homologous points according to the evolutionary phase.
Details about the reduction procedure are reported in the Appendix of \citet{incertezze1}.  Given the accuracy in the observational radius data,  a linear interpolation in time was performed for every reduced track in order to ensure that the separation in radius between consecutive track points in the $10 \sigma$ range from the observational targets was less than one-quarter of the observational radius uncertainty.

\section{Stellar model calibrations}\label{sec:results-AiPhe}

The standard approach of the FRANEC evolutionary code is based on a damping of diffusion velocities in the outermost layers of the stars, but it was not enough to mitigate the drop in the MS, which is around 0.1 dex on average but can reach 0.2 dex. This drop nearly cancels out after the first dredge-up. Therefore, the surface metallicities of the two stars are predicted by stellar models to be significantly different.  
However, the only metallicity constraint available for the system comes from the analysis by \citet{Andersen1988}. These authors measured the metallicity of both stars and detected a spread of 0.04 dex, the more evolved stars having higher surface metallicity. However, the presence of systematic errors and biases suggested a prudential common estimate of $-0.14 \pm 0.1$. 

In light of the theoretical difficulty discussed in Sect.~\ref{sec:intro} to unambiguously predict the surface metallicity of stars in the mass range of Ai Phe before the first dredge-up, 
in the following we investigate two different configurations. In the first fit, we adopt the surface [Fe/H] value resulting from the stellar evolutionary code; in a second fit, we modify it by fixing its value to the initial one. In this second scenario, we still have the effect of the microscopic diffusion on the stellar interior, because we merely block its effect on the surface. 
The choice to fully inhibit the efficiency of the microscopic diffusion allows us to mimic the effect shown by \citet{Moedas2022} in their Fig.~1, that is, a cancellation of the surface metallicity drop owing to the effect of the radiative levitation. While this assumption may be too drastic for a model of 1.2 $M_{\sun}$, nonetheless it sets an extreme reference. The comparison of the calibrations obtained under the two scenarios allows us to investigate their robustness to
different choices of the efficiency of the microscopic diffusion in this critical mass range. It should also be  noted that the adoption of the initial [Fe/H] as an observational constraint is justified for Ai Phe system, because the primary star already experienced the first dredge-up. In a different binary system, with both stars still on the MS, this assumption would be questionable because the initial [Fe/H] value could not safely be assumed as representative of at least one of them.

\subsection{Surface [Fe/H] taking into account diffusion}
Despite the remarkable precision of the observational constraints, the fitting procedure was unable to clearly identify a unique solution, as several acceptable solutions for the system are possible. The kernel density estimator of the marginalised core overshooting parameter $\beta$ (left panel in Fig.~\ref{fig:d-ov-age-AiPhe}) suggests the presence of possible multiple solutions for the system fit. Three of them are located at low or intermediate $\beta$ values up to $\beta \approx 0.14,$ while one is near the upper edge of the explored range. 
The examination of the joint 2D density in the age versus overshooting parameter plane helps us to gain insight into the solution substructure. 
As shown in the right panel of Fig.~\ref{fig:d-ov-age-AiPhe}, the three peaks at low and intermediate overshooting have a similar age, while the peak close to the upper $\beta$ range has a significantly lower age.

\begin{figure*}
	\centering
	\includegraphics[width=8.cm,angle=-90]{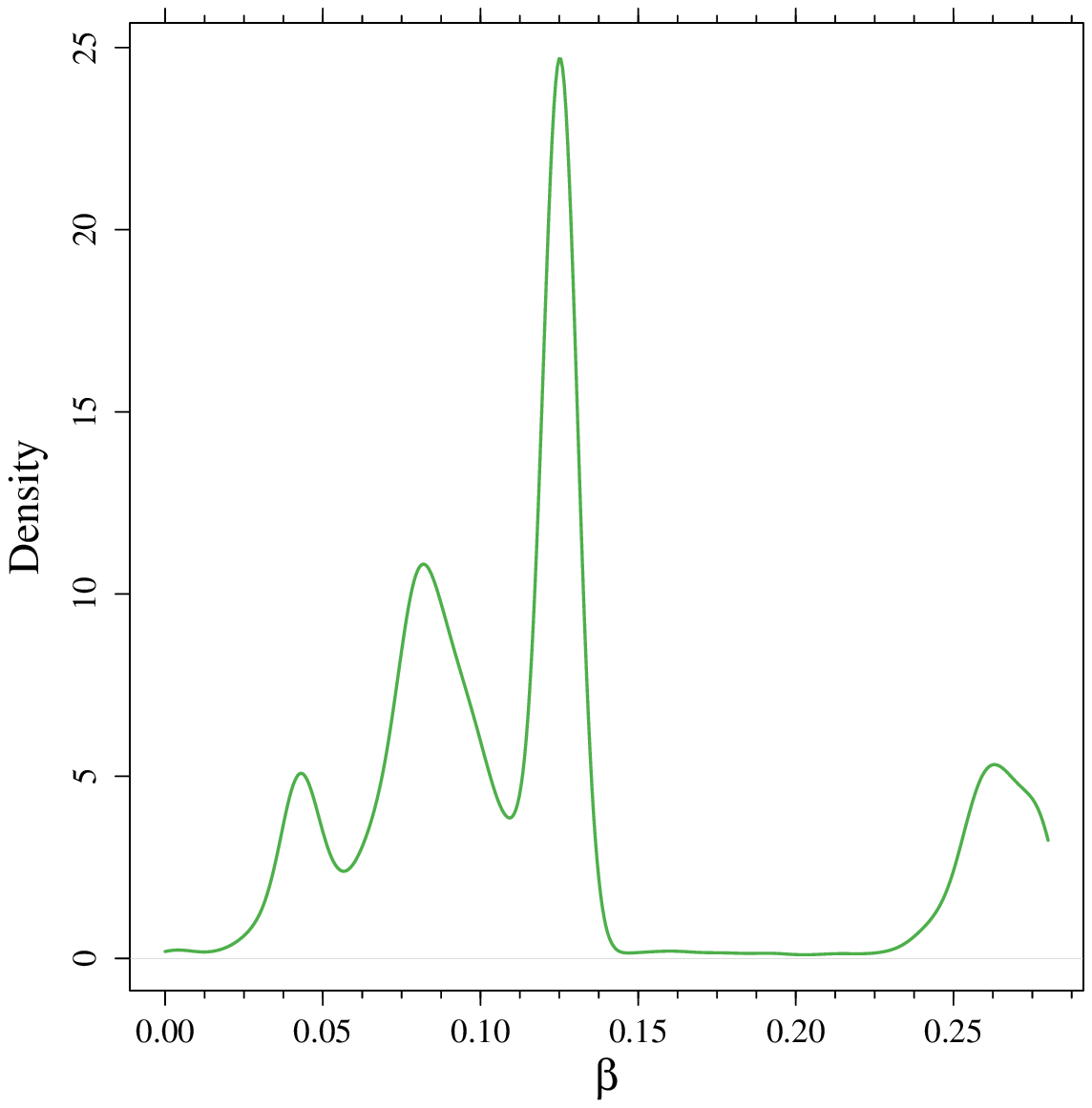} 
	\includegraphics[width=8.cm,angle=-90]{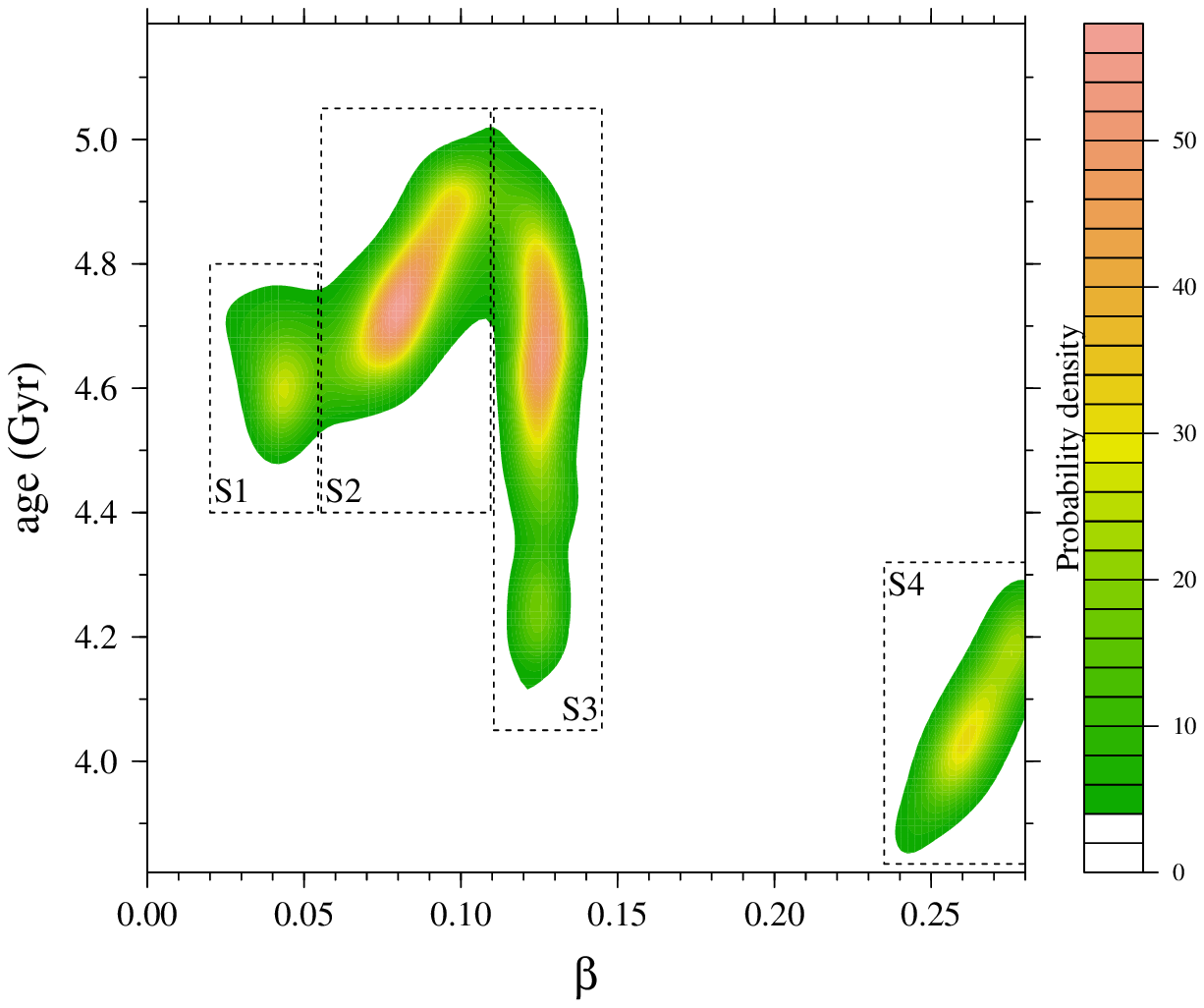}
	
	\caption{Fit of the Ai Phe system with surface [Fe/H] abundances resulting from model calculations with diffusion. {\it Left}: Density of probability for the estimated core overshooting parameter of the Ai Phe system. {\it Right}:  Joint 2D density of probability for the estimated overshooting parameter $\beta$ and the system age.}
	\label{fig:d-ov-age-AiPhe}
\end{figure*}

According to the position of the peaks, we identified four different solutions, labelled S1 to S4 at increasing $\beta$ values, as follows:
\begin{itemize}
        \item S1: solutions with $\beta < 0.055$;
        \item S2: solutions with $0.055 \leq \beta <0.11$;
        \item S3: solutions with $0.11 \leq \beta <0.15$;
        \item S4: solutions with $\beta \geq 0.15$.
\end{itemize} 

The results of the fit, divided according to these regions, are reported in Table~\ref{tab:fit-bin}.
The presence of different islands of solutions stems directly from the 
degeneracy in the impact of the chemical composition and $\beta$  on the stellar age. This effect, which is already discussed in the literature \citep[e.g.][]{Kirkby-Kent2016, TZFor, Constantino2018}, 
prevents us from firmly constraining the $\beta$ value because a set of observational constraints can be reproduced by  different values of the parameters governing the stellar evolution. 

\begin{table*}
	\centering
	\caption{Results of the Ai Phe  binary system fitting in the four identified solution islands with variable surface [Fe/H].}
	\label{tab:fit-bin}
	\begin{tabular}{lcccc}
		\hline\hline
		& S1 & S2& S3 & S4 \\
		\hline 
		$Y$ &  $0.260_{-0.000}^{+0.001}$ & $0.261\pm0.001$ & $0.263_{-0.003}^{+0.016}$ & $0.277 \pm 0.003$\\
		$Z$ &  $0.0115_{-0.0003}^{+0.0006}$ & $0.0123_{-0.0006}^{+0.0008}$ & $0.0124_{-0.0010}^{+0.0018}$ & $0.0096_{-0.0007}^{+0.0008}$ \\
		$\beta$ &  $0.042_{-0.007}^{+0.006}$ & $0.084_{-0.011}^{+0.014}$ & $0.125_{-0.002}^{+0.003}$ & $0.265 \pm 0.012$ \\
		age (Gyr) &   $4.62_{-0.05}^{+0.09}$ & $4.76_{-0.09}^{+0.12}$ & $4.64_{-0.25}^{+0.16}$ & $4.07_{-0.10}^{+0.12}$ \\
		\hline
		\multicolumn{5}{c}{Fit parameters}\\
		\hline
		$T_{\rm eff,1}$ (K) & 5065 & 5034 & 5040 &  5213 \\ 
		$T_{\rm eff,2}$ (K) &  6227 & 6189 & 6249 & 6216 \\ 
		$R_1$ ($R_{\sun}$) &  2.9307 & 2.9301 & 2.9301 & 2.9309\\ 
		$R_2$ ($R_{\sun}$) &  1.8033 & 1.8038 &1.8038 & 1.8029\\ 
		${\rm [Fe/H]}_1$ & -0.08 & -0.04 & -0.04 & -0.14 \\
		${\rm [Fe/H]}_2$ & -0.26 & -0.20 & -0.15 & -0.26\\	
		\hline                
		$\chi^2$ & 3.4 & 2.8 & 3.2 & 7.3 \\ 
		\hline
	\end{tabular}
	\tablefoot{In the first for raw we report: $Y$, initial helium abundance; $Z$, initial metallicity; $\beta$, convective core overshooting parameter; age in Gyr) of the system.} 
\end{table*}

Looking in detail at the proposed solutions, it appears that S4 has a significantly poorer goodness of fit ($\chi^2 = 7.3$) than the others ($\chi^2 \approx 3$). A formal assessment of the goodness of fit is only asymptotically  appropriate when the $\chi^2$ statistic is evaluated over a discrete grid (see \citealt{Frayn2002,goodness2021} for more detail on this topic), approaching the underlying distribution for an infinitely dense grid. However, a rough estimate, assuming 2 degrees of freedom (6 observational constraints and 4 parameters), provides a $P$ value of about 0.02, suggesting that the fit S4 is remarkably poor. Therefore, in the following we restrict the investigation to the solution at low and intermediate $\beta$.

The structure of the solutions is further complicated because S3 is composed of a pool of models. The dominant peak at an age of about 4.7 Gyr corresponds to low-helium models ($\Delta Y/\Delta Z \approx 1.1$), while a secondary, much lower peak is located around 4.2 Gyr. This secondary peak corresponds to a very high helium-to-metal enrichment ratio close to 3.0. However, the relevance of this secondary peak is low, as it accounts for about 12\% of the models in the S3 island. Removing the solutions around the secondary peak increases the S3 best-fit age to $4.67\pm 0.15$ Gyr, with little modification of the S3 initial chemical abundances.  

\begin{figure*}
        \centering
        \includegraphics[width=6.2cm,angle=-90]{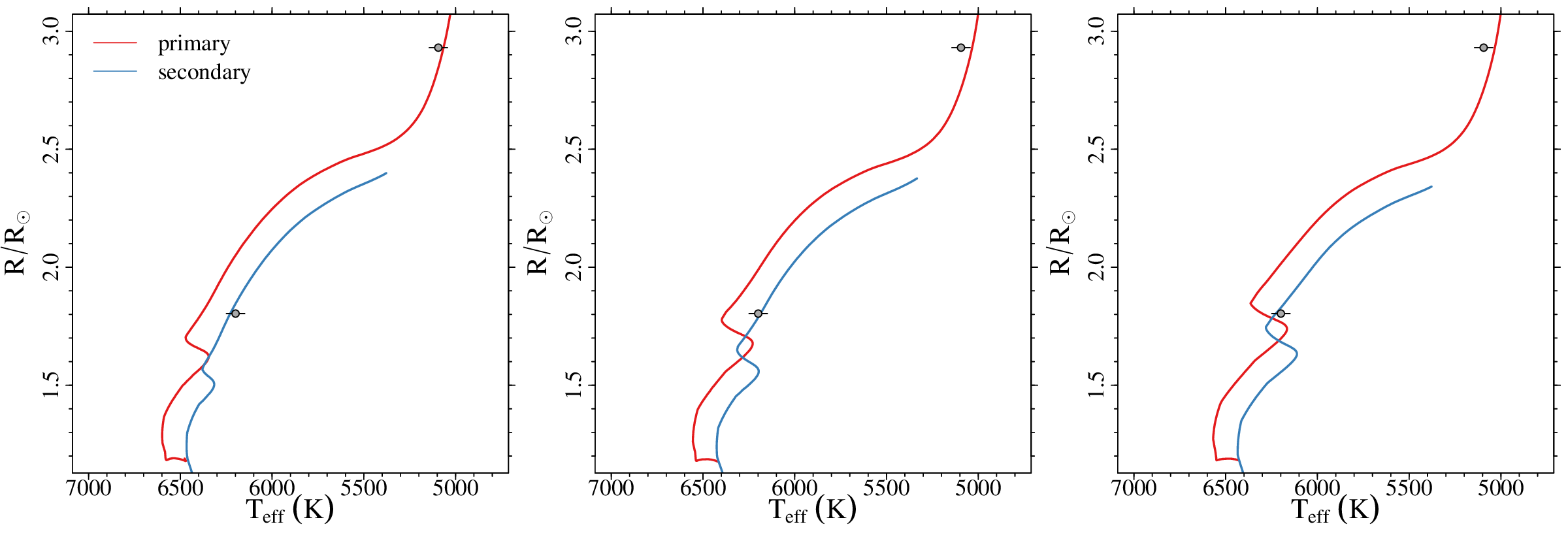} 
        \caption{Comparison between the observational values of effective temperature and radius of the two stars (grey circle) and the evolutionary tracks for the different solutions found in the analysis. {\it Left}: Solution S1.  The error bars correspond to 1 $\sigma$ errors, and the errors in radii are too small to see.
        {\it Middle}: Same as in the left panel but for solution S2. 
        {\it Right}: Same as in the left panel but for solution S3.     
        }
        \label{fig:HR-AiPhe}
\end{figure*}

As expected, the effect of microscopic diffusion is clearly shown in Table~\ref{tab:fit-bin}, which reports a difference of 0.10 to 0.20 dex in surface [Fe/H] fitted values. The large error in the [Fe/H] constraints mitigates the problem, allowing the pipeline to find solutions at different but compatible surface metallicities.

The agreement between observational data and best-fit evolutionary tracks is displayed in  Fig.~\ref{fig:HR-AiPhe} in the radius versus effective temperature plane. The corresponding best-fit models are not shown in order to improve readability, but correspond to the points where the radius assumes the observational value. The identified evolutionary stages show a primary ascending the RGB and a secondary in the SGB, in agreement with the analysis of \citet{Kirkby-Kent2016}. The evolutionary stage of the secondary is slightly different among S1, S2, and S3, progressively moving towards an earlier SGB phase.
Overall, the pooled age estimate from S1 to S3 solutions is $4.70_{-0.14}^{+0.13}$ Gyr when considering the secondary peak in the S3 island. Neglecting the secondary solution in the S3 basin has a negligible impact, only modifying the age estimate by 0.01 Gyr. 

The proposed fits show a clear preference for low-helium models; more precisely, a share of about 75\% of the solution in S1 to S3 areas are at $\Delta Y/\Delta Z < 1.2$. 
This result is at odds with a recent investigation
performed on the Hyades cluster by \citet{Tognelli2021}, who obtained a value of $\Delta Y/\Delta Z = 2.03 \pm 0.15$. For comparison, only 4\% of the Ai Phe system solutions lie in the $1 \sigma$ range [1.87, 2.18].
We performed a direct test by restricting the fitting grid to a $2 \sigma$ range [1.7, 2.3] around $\Delta Y/\Delta Z = 2.0$. In this scenario, the algorithm finds a solution for only 8\% of the Monte Carlo experiments. The solution is strongly peaked at $\beta = 0.125 \pm 0.005$ and an age of $4.56 \pm 0.12$ Gyr in the S3 island. The $\chi^2$ of the solution is 4.1, suggesting an acceptable agreement with data. However, as discussed above, better solutions (according to the goodness-of-fit statistic) exist for the non-restricted grid. 

Interestingly, the result for the Ai Phe fit agrees with the findings of \citet{TZFor} for the TZ For binary system, a system with an evolved primary star ---already in the central He burning phase---  and a secondary close to the hydrogen depletion. The fit of that system, performed using an identical pipeline and stellar tracks computed with the same stellar evolutionary code, resulted in $\Delta Y/\Delta Z$ of close to 1.0, ruling out solutions at higher helium-to-metal enrichment ratios.

\subsection{Surface [Fe/H] fixed at the original value}

The solutions proposed by the fit performed whilst imposing that surface [Fe/H] abundance be fixed at the initial value, as in the case of inefficient miscroscopic diffusion, show some differences with respect to those discussed in the previous paragraph. Figure~\ref{fig:HR-AiPhe-nofeh} shows the presence of two islands, identified as follows:
\begin{itemize}
        \item F1: Solutions with $\beta < 0.065$ and an age of less than 4.75 Gyr. A long tail extends at higher age but it is not considered a part of this island.
        \item F2: Solutions with  $\beta \geq 0.15$.
\end{itemize}
The corresponding best-fit values and the estimated parameters are collected in Table~\ref{tab:fit-bin-nofeh}.

\begin{table}
        \centering
        \caption{Results of the Ai Phe  binary system fitting with fixed surface [Fe/H].}
        \label{tab:fit-bin-nofeh}
        \begin{tabular}{lcc}
                \hline\hline
                & F1 & F2 \\
                \hline 
                $Y$ & $0.260_{-0.000}^{+0.001}$ & $0.274_{-0.007}^{+0.004}$\\
                $Z$ &  $0.0114_{-0.0003}^{+0.0003}$ & $0.0093_{-0.0007}^{+0.0005}$\\
                $\beta$ & $0.040 \pm 0.004$ &  $0.265_{-0.015}^{+0.012}$\\
                age (Gyr) &  $4.58_{-0.04}^{+0.05}$ &  $4.09 \pm 0.12$ \\
                \hline
                \multicolumn{3}{c}{Fit parameters}\\
                \hline
                $T_{\rm eff,1}$ (K) & 5072 & 5224 \\ 
                $T_{\rm eff,2}$ (K) & 6237 & 6226 \\ 
                $R_1$ ($R_{\sun}$) & 2.9301 & 2.9309 \\ 
                $R_2$ ($R_{\sun}$) & 1.8040 & 1.8029 \\ 
                ${\rm [Fe/H]}_1$ &   -0.07  & -0.14\\
                ${\rm [Fe/H]}_2$ &  -0.07 &  -0.14\\    
                \hline                
                $\chi^2$ & 1.9  &  7.2 \\ 
                \hline
        \end{tabular}
\end{table}

It is apparent that the F1 solution corresponds to S1. As opposed to the previously discussed solutions, the two prominent S2 and S3 islands disappear, the only remnant of S2 being the long tail towards higher age stemming from F1. Interestingly, there is no remnant of the S3 solution at $\beta \approx 0.12$. 
Solution F2 corresponds to S4. Similarly to the previous scenario, this solution has a relatively high $\chi^2$ and can therefore be disregarded as a system fit.

Including the tail in the F1 island (see Fig.~\ref{fig:HR-AiPhe-nofeh}) modifies the estimated age of the system to $4.62_{-0.06}^{+0.13}$ Gyr, which is only 2\% younger than the age estimated before. The corresponding convective core overshooting parameter is $\beta = 0.042_{-0.005}^{+0.027}$.  
      
Similarly to the previous results, in this case low-helium models are  also preferred. As in the previous section, we directly tested a restriction of the grid in the $2 \sigma$ range around $\Delta Y/\Delta Z = 2.0$. This change  has drastic consequences for the results: only about 1\% of the Monte Carlo experiments converged towards a bimodal solution, that is, at $\beta = 0.032_{-0.008}^{+0.030}$ and at an age of $4.59_{-0.04}^{+0.12}$ Gyr (F1 island) and $\beta = 0.276_{-0.054}^{+0.003}$ and an age of $4.22_{-0.06}^{+0.04}$ Gyr (F2). As opposed to the previous section, the $\chi^2$ of these solutions are relatively high, at 11.2 and 9.9, respectively, suggesting very poor fits, and confirming the difficulties  the algorithm has in providing solutions in this scenario.

\begin{figure*}
        \centering
        \includegraphics[width=8cm,angle=-90]{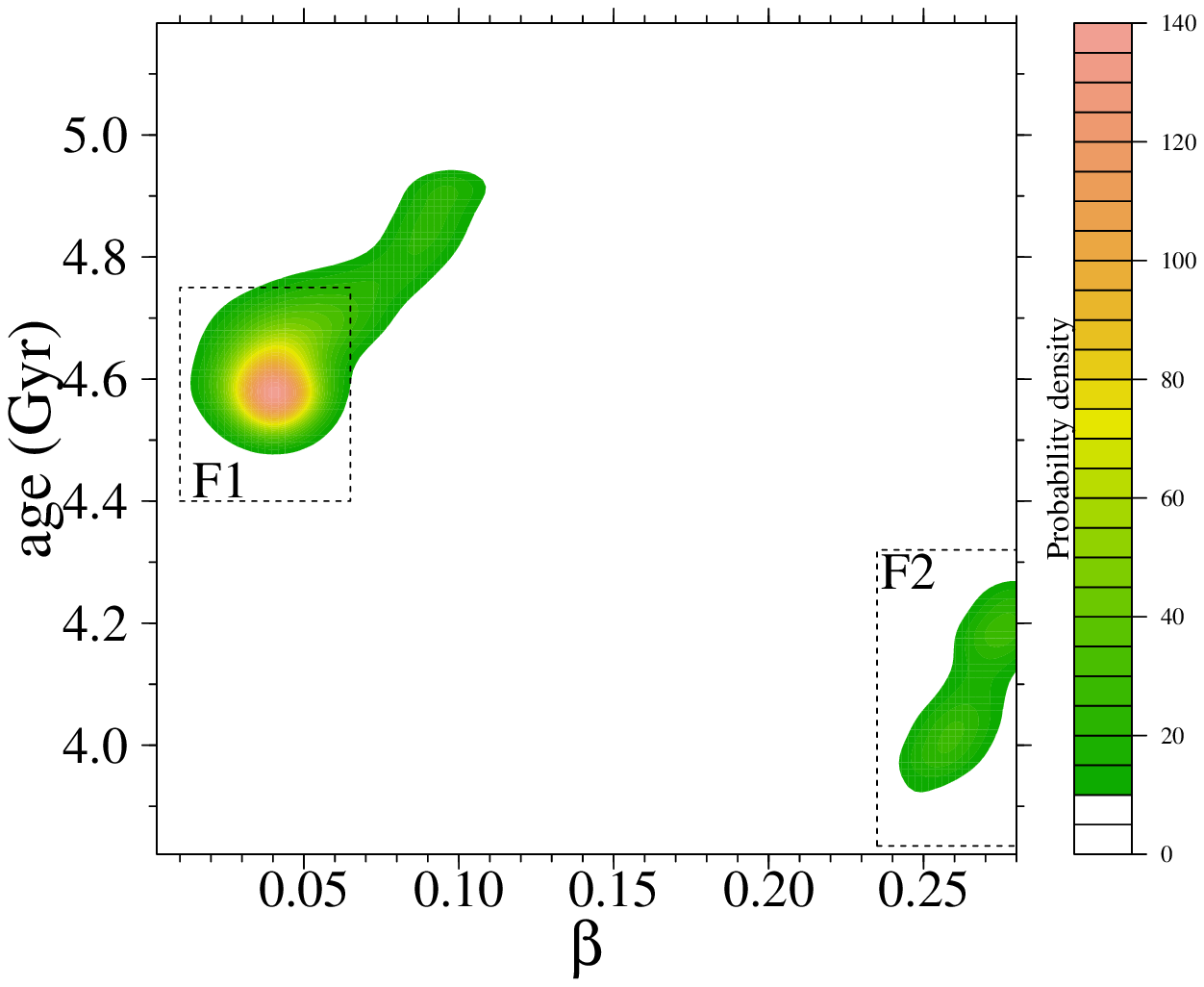} 
        \includegraphics[width=8cm,angle=-90]{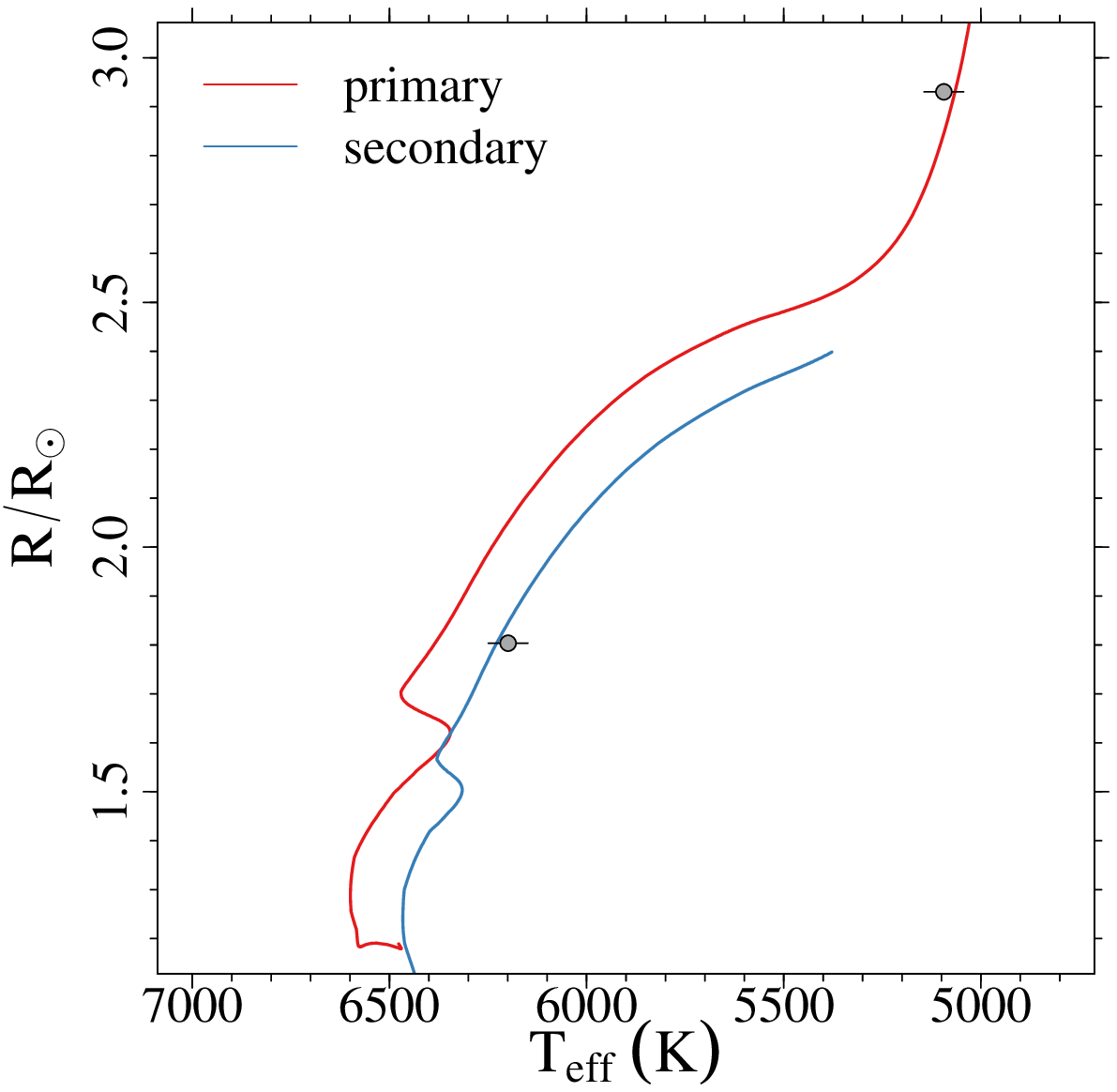} 
        \caption{Fit of the Ai Phe system with fixed surface [Fe/H] abundance. ({\it Left}): Joint 2D probability density in the $\beta$ vs. age plane. ({\it Right}): Comparison between the observational values of effective temperature and radius of the two stars (grey circle) and the evolutionary tracks for the solution F1 found in the analysis.  The error bars correspond to 1 $\sigma$ errors.   
        }
        \label{fig:HR-AiPhe-nofeh}
\end{figure*}

\section{Discussion}\label{sec:discuss}

Despite the extraordinary precision in observational constraints, we were unable to obtain an undisputed fit of the system. This indetermination, which possibly limits the calibrating power of some binary systems, has already been recognised and discussed in the literature \citep[e.g.][]{TZFor, Constantino2018, Johnston2019}.

The system fits under the two explored scenarios show some similarity, but also several significant differences. First of all, we find the estimated age of Ai Phe to be very robust to different assumptions about the variation of [Fe/H] during stellar evolution. The improved precision in the stellar radii and effective temperatures allowed us to reach a 3\% precision in the system age.
However, this reassuring finding cannot be taken as a general rule, because systems composed of stars of different masses or in different evolutionary phases may react differently to changes in the surface [Fe/H].     
On the other hand, a sharp calibration of the core overshooting parameter was not achievable. The differences between the obtained calibrations under the two scenarios are quite relevant. In the scenario with unmodified surface [Fe/H], the degeneracy between $\beta$ and the initial chemical composition allows solutions in a wide range of $\beta$. The estimate is barely constrained, and found to be in the  range of 0.04 to 0.12. When the surface metallicity is kept fixed at its original value, the pipeline points towards a cleaner solution in the range of 0.03 to 0.07. The two dominant solutions around $\beta$ from 0.07 to 0.12 disappear. While none of these fits can be considered  as a gold standard, the differences obtained by only modifying the variation of surface [Fe/H] during stellar evolution ---a parameter that is currently far from fully understood--- are striking. As such this investigation raises some doubt as to the robustness of the calibrations obtained in the mass range between 1.2 $M_{\sun}$ and 1.5 $M_{\sun}$.

Thanks to the availability of high-quality observations, this system has been extensively investigated in the literature, with the most recent investigation performed by \citet{Kirkby-Kent2016}.
Relying on WASP photometry, which allowed accuracy on the  radius of about 1\% and effective temperature uncertainty of about 150 K, the authors suggested an age of $4.39 \pm 0.32$ Gyr for an initial helium value of $0.26_{-0.01}^{+0.02}$. Even after the relevant improvement in the observation accuracy, these values are in agreement with those reported here at the $1 \sigma$ level.

Regarding the convective core overshooting efficiency, relying on older and less accurate observations, \citet{Claret2016} and \citet{Claret2017} performed a calibration of the overshooting parameter working with both step and diffusive overshooting approaches. These authors obtain values of $\beta = 0.04$ and 0.00 for the primary and secondary star, respectively, in the first scenario, and $f_{\rm ov} = 0.00$ in the second scenario at an age of 4.38 Gyr.
Many differences exist between the adopted fit frameworks, input physics in the stellar track computation, and observational constraints, making a direct comparison with those results questionable. Moreover, as discussed by \citet{BinTeo}, a comparison of overshooting parameters without considering the overshooting scheme implemented in the evolutionary codes should be avoided. A safer comparison is between physical quantities, such as the convective core mass, whenever possible. This is not the case for the Ai Phe system, because the secondary is in an evolutionary stage where the convective core has already vanished.

While the pipeline was able to produce a satisfactory fit of the system, there are nonetheless some possible distortions and systematic effects that are worth discussing. First of all, the fit was performed at fixed input physics. It is well known that different pipelines, adopting different estimation algorithms and different stellar tracks, might obtain different estimates of the system fundamental parameters  \citep{Reese2016, Stancliffe2016,  SilvaAguirre2017, TZFor, Gallenne2023}. A conservative estimation of the precision achievable on the age of Ai Phe is probably double the figure obtained by a single pipeline approach, that is, about 7\%-10\%.   
 
We performed the fit assuming identical initial chemical composition and a common age. These assumptions are easily justified because the formation scenario supposes both stars formed nearly simultaneously from the same matter.  
Some other assumptions are more questionable. A common solar-scaled mixing length value was adopted for both stars and this values was kept fixed during the evolution. \citet{Trampedach2013} investigated the mixing length 
variation during stellar evolution using radiation hydrodynamics simulations. As discussed by \citet{Kirkby-Kent2016}, this effect might cause the primary star in the Ai Phe system to have a slightly larger mixing-length parameter value. However, this effect was considered to have little impact on the estimated age. 
Some authors \citep[e.g.][]{Claret2007, Graczyk2016} deal with this problem by introducing further degrees of freedom in the fit, allowing the two stars to have different mixing-length values. However, this approach may lead to over-fitting the system and, something that is more problematic, may mask possible mismatches between stellar tracks and real observations by only adjusting the mixing length.
While this fine tuning may be necessary to obtain a reliable fit of a system \citep[see the discussion in][]{Gallenne2023}, it is possibly only a `cosmetic' remedy. Masking the existing impossibility to fit a system under canonical assumptions and 
hiding the difficulties by allowing every star to have its own mixing-length value does not contribute to developing more sensible stellar models. 

Identical considerations apply to the choice of having a common core-overshooting parameter value for both stars. Ultimately, allowing the stars to have independent overshooting efficiencies or mixing-length values would lead to additional free parameters, which would reflect not only the additional mixing, but also other effects from any given source, which would remain hidden (see the discussion in \citealt{TZFor} for greater details on this topic). Moreover, while a difference in the overshooting parameter can be theoretically required and justified when the two stars have different masses, it is somewhat questionable for Ai Phe, given that the two stars have nearly identical masses.

\section{Conclusions}\label{sec:conclusions}

Taking advantage of recently available, very precise observational data  for the Ai Phe double-lined eclipsing binary system \citep{Miller2020}, we attempted to constrain the age and the efficiency of the convective core overshooting of the two stars under different assumptions. 
To do this, we used the SCEPtER pipeline \citep{scepter1, eta, binary} on a dense grid of stellar models computed ad hoc.

We were able to obtain a satisfactory but multi-modal fit for the system at   
age $4.70^{+0.13}_{-0.14}$ Gyr, with an overshooting parameter $\beta$ in the range of 0.04--0.12. The estimated age was in agreement with the results of \citet{Kirkby-Kent2016}, who suggested an age of $4.39 \pm 0.32$ Gyr.
The fitting grid of the stellar track adopted for these estimates was computed including the effect of microscopic diffusion, which alters the surface metallicity [Fe/H] during stellar evolution, which impacts the fit of the system. 
Due to the fact that the efficiency of microscopic diffusion in the mass range of the system ---around 1.2 $M_{\sun}$--- is still debated, we tested an alternative scenario by blocking the update of the surface metallicity, but allowing microscopic diffusion in the interior layers. 
This test is quite relevant because the two stars are in different evolutionary phases: while the primary has already experienced the first dredge-up, almost recovering its initial surface metallicity, the secondary has not. 
The age fitted in this second scenario, of namely $4.62^{+0.13}_{-0.06}$ Gyr, agrees well with the age from the former fit. The most relevant difference is in the convective core overshooting calibration, because this scenario points towards a sharp solution in the 0.03 to 0.07 range. 

The comparison of the two solutions provides satisfactory confirmation of the robustness of the age estimates obtained by our pipeline. The same conclusion was obtained for the CPD-54 810 binary system \citep{CPD2023}. On the other hand, it suggests great care should be taken when adopting binary systems for parameter calibrations, because the obtained parameters may only reflect the decisions by the modellers in their stellar model computations \citep[see e.g.][]{Constantino2018, Johnston2019}. In this specific case, precise measurements of individual surface [Fe/H] would be of utmost importance in helping us to judge which scenario is the most reliable. While this is relevant for the Ai Phe system, given the evolutionary phases of the two stars, it may not be so for a different system. As already discussed in previous works \citep{TZFor, CPD2023}, every system provides different challenges and poses interesting questions about the possibility to use them to constrain free parameters in stellar model computations.

\begin{acknowledgements}
We thank our anonymous referee for the very useful comments and suggestions.
G.V., P.G.P.M. and S.D. acknowledge INFN (Iniziativa specifica TAsP).
\end{acknowledgements}

\bibliographystyle{aa}
\bibliography{biblio}

\end{document}